\documentclass[11pt,preprint]{aastex}

\slugcomment{To appear in May 20, 2004 issue of ApJ}

\shortauthors{Nakajima et al.}
\shorttitle{L and T dwarfs}

\begin{document}

\title{Spectral Classification and Effective Temperatures of L and T Dwarfs
Based on Near-Infrared Spectra  \altaffilmark{1}}

\author{Tadashi Nakajima\altaffilmark{2}}
\affil{National Astronomical Observatory, 2-21-1, Osawa, Mitaka
181-8588, Japan}
\email{tadashi.nakajima@nao.ac.jp}

\author{Takashi Tsuji}
\affil{Institute of Astronomy, The University of Tokyo,
 2-21-1, Osawa, Mitaka
181-0015, Japan}
\email{ttsuji@ioa.s.u-tokyo.ac.jp}

\and

\author{Kenshi Yanagisawa}
\affil{Okayama Astrophysical Observatory, Kamogata, Okayama 719-0231}
\email{yanagi@oao.nao.ac.jp}


\altaffiltext{1}{Based on the data collected at the Subaru Telescope, 
which is operated by the National Astronomical Observatory of Japan}


\begin{abstract}
We have obtained near-infrared
spectra of L dwarfs, L/T transition objects
and T dwarfs using Subaru.
The resulting spectra are examined in detail to see their dependence 
on the spectral types. One question is where the methane feature
appears: We suggest that it appears at L8 and marginally at L6.5.
The water bands at 1.1 and 1.4 $\mu$m do not necessarily show steady
increase towards later L types but may show inversion at late L types.
This does not necessarily imply that the spectral types do not 
represent a temperature sequence, but can be interpreted as due to
compensation of the increasing water abundance by the 
heavier of dust extinction
in the later L types. We confirm that the FeH 0.99 $\mu$m bands appear 
not only in the late L dwarfs but also in the early T dwarfs. We suggest
that FeH could be dredged up by the surface convective zone induced by 
the steep temperature gradient due to the large opacity of the dust
cloud itself and will be replenished constantly by the convection.

We have obtained bolometric luminosities of the objects with known
parallaxes in our sample, first
by integrating the spectra between 0.87 and 2.5 $\mu$m,
and second  by the $K$-band bolometric correction. 
Apart from an L3 dwarf, the bolometric luminosities obtained by
both methods agree well and this implies that the $K$-band 
bolometric correction, which is  obtained by 
the use of the Unified Cloudy Models,
can be applied to obtain the bolometric luminosities and
effective temperatures of the L and T dwarfs
with known parallaxes from the literature. 
The relation between the effective temperature and spectral type
derived from the $K$-band bolometric correction
shows  monotonic 
behavior throughout the L-T sequence.

\end{abstract}




%

\section{INTRODUCTION}

The very coolest stars and the warmer brown dwarfs require a new
spectral class, known as `L' \citep{mar97,kir99}. The L dwarf
sequence is characterized by the disappearance of the red TiO and VO
bands from the optical (0.6$-$1.0 $\mu$m) spectrum, the increasing
dominance at those wavelengths by broad absorption resonance lines of
Na I and K I, and strong H$_2$O absorption bands and persistent CO
overtone bands in the 1$-$2.5 $\mu$m region
\citep{mar99,kir99,kir00,leg01,rei01b}.
In terms of broadband colors, L dwarfs are characterized by very red
infrared colors (e.g. $J-K>1.3$).
Even cooler brown dwarfs require the additional class, `T' \citep{kir99}.
In T dwarfs, the CO bands are replaced by stronger and more extensive
absorptions of CH$_4$ in the H and K bands, and there is further
strengthening of water bands \citep{opp95,geb96,str99,bur99}. 
T dwarfs are characterized 
by blue infrared colors (e.g. $J-K \sim 0$).

Spectral classification of L and T dwarfs including L/T transition
objects has been made by \citet{geb02} and 
\citet{bur02a}. L dwarfs
are now classified from L0 to L9 and T dwarfs from T0 to T8.
The classification scheme is purely observational and uses indices
in the 1$-$2.5 $\mu$m region
related to H$_2$O, CO, CH$_4$ and near-infrared colors.
We reexamine the spectral classification using the spectra obtained at 
Subaru. We pay special attention to the behavior of FeH and
the features related to  H$_2$O and CH$_4$.  

Once the classification scheme is  established,
the next step is to find the correspondence between  
effective temperatures and spectral type and elucidate the physical meaning of 
the spectral classification.
Some authors have derived relations between the 
effective temperatures and spectral type
for L dwarfs \citep{leg02,dah02} and for L and T dwarfs
\citep{bur01}, but there
has not been a work in which the relation was obtained from
effective temperatures based on 
bolometric correction derived from photospheric models which are
applicable throughout  the L-T sequence.

Some of the objects in our sample have known parallaxes.
We obtain bolometric luminosities and effective temperatures
for these objects by integrating the observed spectra between 
0.87 and 2.5 $\mu$m and by the $K$-band bolometric correction
using the Unified Cloudy Models (UCM) \citep{tsu02,tsu03a}.
The comparison of the two methods implies that the $K$-band bolometric 
correction leads to reasonable effective temperatures except for early 
L type. Encouraged by this analysis, we apply the $K$-band bolometric
correction to the objects whose parallaxes are known from the
literature \citep{bur01,dah02,tin03}.

The paper is organized as follows. In \S2, observations,
data reduction and the sample for the analysis
are described. Spectral classification 
is discussed and identification of
spectral features is reexamined in \S3. 
We confirm the validity of the $K$-band bolometric correction derived
by UCMs with the integrated fluxes as noted above. We then obtain
bolometric luminosities of the objects with known parallaxes and
the relation  between
the effective temperature and spectral type in \S4.
The summary of the paper is given in \S5.

\section{OBSERVED SPECTRA}

\subsection{Observations}

Observations were made on 2002, June 3 UT at the Subaru telescope
using imaging and grism modes of CISCO \citep{mot02}.

We obtained spectra of six objects (Table 2), for four of which 
$JHK^\prime$ photometry was
also obtained (Table 1). The six objects were selected to sample
the L-T sequence widely from the objects in \citet{geb02} and \citet{bur02a}.

The sky was clear and the seeing was less than $0.\hspace{-2pt}''5$
at 2 $\mu$m throughout the night.
CISCO employs the MKO-NIR filters \citep{sim02,tok02}.
Photometric observations were made when the objects were near transit,
and the airmasses were less than 1.2. For photometric
calibration, UKIRT faint standards \citep{haw01} were used. 

In the case of spectroscopy,
the slit width of 
$0.\hspace{-2pt}''5$ was sampled at $0.\hspace{-2pt}''105$
pixel$^{-1}$ and the resolutions of $zJ$ (0.882-1.400 $\mu$m),
$JH$ (1.056-1.816  $\mu$m) and wide-$K$ (1.850-2.512 $\mu$m) grisms were
550, 400 and 600 respectively. The targets were nodded along the slit,
and observations taken in ABBA sequence, where A and B stand for the
first and second positions on the slit.
Integration time is summarized in Table 2.
SDSS2249+00 was observed at the end of the night and only a $JH$ spectrum was 
obtained for it.
Spectra of a nearby F/G star were obtained before or after
the observations of each object.

\subsection{Data Reduction}

For photometric data
reduction, aperture photometry was made using the APPHOT package in
IRAF. Photon statistics gave uncertainties much less than 0.05 mag. However,
we estimated that airmass differences between the targets and standards
and the difference between the UKIRT and MKO-NIR systems introduced
uncertainties of about 0.05 mag.

Spectral data were reduced using IRAF. 
The effect of the terrestrial atmosphere was removed by dividing
by the spectra of nearby F/G stars after removing hydrogen
recombination
lines seen in their spectra.
Spectral segments were individually flux calibrated using the 
photometric data given in Table 1. Each segment and  
spectral data of Vega \citep{ber95} were integrated over
the appropriate filter profile.
Vega was assumed to be zero magnitude at all wavelengths,
and the target flux was scaled to match the broadband photometry.

\subsection{Sample for Analysis}

In addition to the six objects for which spectra were obtained in June 
2002, we added three objects whose spectra were
obtained previously
by IRCS \citep{kob00} on Subaru, 2MASS1146+22, 2MASS0920+35, and
2MASS1507$-$16 
from a previously published paper
\citep{nak01} and Gl229B from the literature
\citep{geb96,leg99}. There are in total ten objects as listed
in Table 3. 
The sample ranges in spectral type from L3 to T8.  For six of the
ten objects
trigonometric parallaxes are known 
from the literature (Burgasser 2001; Dahn et al. 2002; 
Tinney et al. 2003; references therein): They are
2MASS1146+22, 2MASS1507$-$16, 2MASS1523+30, SDSS1254$-$01, 
Gl229B, and 2MASS1217$-$03. 

2MASS1146+22 is a close binary \citep{rei01a} whose spectral type is L3 
\citep{kir99}. The observed magnitude difference is small at $I$ (0.3 mag)
and Reid et al. estimate that the difference in bolometric
magnitude is 0.15.

2MASS0920+35 is also a close binary whose spectral type is L6.5
\citep{rei01a,kir99}. Weak methane absorption features are seen
at $H$ and $K$ bands \citep{nak01}.
The magnitude difference at $I$ is 0.44 and Reid et al.  estimate
that the difference in bolometric magnitude is 0.15.
The parallax of this object is not known and we use this object 
only for the discussion of its spectrum.

\section{NEAR-INFRARED SPECTRA AND SPECTRAL CLASSIFICATION}

We examine our spectra which cover a representative sample of L and T
types in some detail. It turns out that some of the prominent
spectral features remain unidentified and the interpretation of the
identified features is by no means clear yet. In this section, we apply  
the predicted spectral line intensities based on the UCMs discussed 
in a separate paper \citep{tsu03}
as a guide to interpret the observed spectra.

\subsection{The $K$-Band Region}

The spectra of eight objects in the $K$-band region are shown on the 
$\log f_\nu$ scale in Fig.1. The prominent features are CO first
overtone bands at 2.3 $\mu$m in L dwarfs and can be traced up to 
T2 or T3.5 dwarfs in our sample. The methane bands at 2.2 $\mu$m are
quite strong in T dwarfs, and a question is if they are already
seen in late L dwarfs. A very weak bandhead feature may be seen at 2.2 
$\mu$m in the L6.5 dwarf 2MASS0920+35 as already noted previously
\citep{nak01}. The spectrum of another L6.5 dwarf 2MASS1711+22 is a bit
noisy and it is difficult to identify the methane 2.2 $\mu$m bands. In 
L8 dwarf 2MASS1523+30, the presence of the methane 2.2 $\mu$m bands was
previously suggested by \citet{mcl01}. The S/N ratio of our
spectrum of 2MASS1523+30 may be somewhat better than that of McLean et
al. and the methane 2.2 $\mu$m bands can be clearly seen in Fig.1.
Thus, the methane 2.2 $\mu$m bands can be deemed as detected at L8.
This better S/N ratio is probably due to the lower spectral
resolution and higher throughput of CISCO on Subaru than NIRSPEC on Keck.

\subsection{The $H$-Band Region}
  
In the $H$-band region, absorption features  are clearly seen at 1.58, 1.59, 
1.61, and 1.625 $\mu$m in L3 and L5 dwarfs as shown by the filled triangles
in Fig. 2. 
Of these features, those at 1.58, 1.613, and 1.627 $\mu$m were 
noted in L dwarfs by \citet{rei01b}. On the other hand,  the features 
at 1.583, 1.591, and 1.625 $\mu$m were identified as due to the FeH 
$E^4\Pi - A^4\Pi$ system in the spectra of sunspot as well as of M - L 
dwarfs by \citet{wal01}, who also remarked that the feature 
at 1.611 $\mu$m had a confused appearance.
More recently, the analysis of the FeH bands in M and L dwarfs has been
extended by \citet{cus03}, who showed that the 1.625 $\mu$m
feature is the strongest among the four features prominent in the
$H$-band region. Thus all the four features (1.583, 1.591, 1.611, 
and 1.625 $\mu$m) in our spectra of L3 - L5 dwarfs can be identified with
FeH. 

A question is if the same identification can be applied to the L6.5 and 
L8 dwarfs in our sample. Another possibility is that the 1.63 as well
as 1.67 $\mu$m feature in the L6.5 and L8 dwarfs is due to methane
bands. We have already attributed these features as due to methane in
the L6.5 dwarf 2MASS0920+35 \citep{nak01}, but the possible presence of the 
FeH bands was not known at that time.  The 1.63 $\mu$m feature can also 
be seen in all the L dwarfs  while the 1.67 $\mu$m feature is not clear 
in the L3 - L5 dwarfs and appears first in L6.5. 
Also, the 1.583 and 1.591 $\mu$m features are rather weak in the
L6.5 dwarfs. These observations may support the appearance of methane at
L6.5, and it is at least possible that FeH and CH$_4$ both contribute
to the 1.63 and 1.67 $\mu$m features in L6.5 dwarfs.  
To settle this problem, higher resolution works are needed.
As for 2MASS1523+30, absence of the 1.61 $\mu$m feature suggests
that the 1.63 $\mu$m feature may be due to CH$_4$ rather than FeH. 
The methane features at 1.63 and 1.67 $\mu$m
are clearly seen in the T2 dwarf and strengthen rapidly toward the later
T dwarfs. Where methane bands appear is a critical issue in the spectral
classification of brown dwarfs, since methane is a sensitive indicator 
of temperature. Although the methane bands depend on gravity as well as on
temperature, the increased abundance of methane at the higher gravity
is compensated for to some extent by the increased  H$_2$ CIA which also 
increases with gravity \citep{tsu03}.

It is to be noted that the $H$-band region is by no means a good
continuum window, but is contaminated by molecular bands of unknown
origin in addition to FeH especially in L dwarfs. Also, we notice a
large difference in the spectra of the same spectral type, namely
between the L dwarfs 2MASS1507$-$16 and SDSS2249+00: The $H$ band spectrum of
2MASS1507$-$16 is rather flat 
as are those of the other L dwarfs while that of SDSS2249+00
shows a prominent peaking centered at about 1.7 $\mu$m. The reason for
this difference is unknown at present.

\subsection{The Region between $J$- and $H$- Bands}

In Fig.3, we show the region between $J$ and $H$ bands. It is to be
noted that water bands at 1.4 $\mu$m can be well measured in the five
spectra observed by CISCO despite the water bands in the Earth's
atmosphere. Unfortunately, only the edge of the 1.4 $\mu$m bands, arising in
hotter environments of these cool dwarfs than the Earth's atmosphere,
can be measured in the three objects observed by IRCS.
One interesting result is that the H$_2$O 1.4 $\mu$m band does not appear
stronger  in the L8 dwarf 2MASS1523+30 than in the L6.5 dwarfs 2MASS1711+22 and
2MASS0920+35\footnote{The depression of the edge near 1.32\,$\mu$m is nearly
comparable between 2M1711 and 2M1523 but the overall depression between
1.31 and 1.60\,$\mu$m may be slightly larger in 2M1711 than in 2M1523.}. 
Further, in the 1.1 $\mu$m bands of water, the absorption is clearly
weaker in the L8 dwarf 2M1523 than in L6.5 dwarf 2M1711 as can been seen 
in Fig.4. The spectral types are based on \citet{kir00} and those by 
\citet{geb02} also show the similar types. This result is
somewhat surprising, since the 1.1 and 1.4 $\mu$m water bands were
used as classification criteria by \citet{geb02}.  The water
band strengths, however,  do not necessarily increase linearly with
the L types, and some water bands show little change in their
strengths between L3 and L8 \citep{rei01b,tes01}. 

Although there appears a plateau in the $T_{\rm eff}$ scale between 
L6 and L8 (\S4.2), we think that the spectral types basically 
correspond to a temperature sequence, 
since the present spectral types are determined by the several temperature 
sensitive spectral features in the optical  region on purely empirical 
basis \citep{kir99} and are
consistent mostly with the infrared spectral indices including methane 
bands \citep{bur02a,geb02}. Thus, it is possible that water
bands are weaker in the cooler L dwarfs than in the warmer objects,
while the methane feature at 2.2 $\mu$m is stronger in the L8 dwarf
than in the L6.5 dwarfs as shown in Fig. 1. We
interpret this result as follows. The water abundance should be larger 
in cooler objects, but extinction by  dust clouds, still located in the 
optically thin region in L dwarfs, should be larger in cooler objects
as well. Then the increasing water abundance and the increasing dust
extinction may cancel out in L dwarfs. For this reason, water band
strengths do not show a large increase in L dwarfs \citep{tsu02}, and 
it is no wonder that they show an inversion somewhere in L types. 
The methane feature at 2.2 $\mu$m does not necessarily suffer the same 
effect, either because the dust extinction is already not so
 important in the 2.2 $\mu$m region than the 1.4 $\mu$m region, as also
can be seen in the cloud models by \citet{ack01}, or else because of the more
rapid increase of the methane band strength with the decreasing
effective temperature than in the case of the water bands. We have
confirmed these results by a detailed computation of the spectral line
intensities based on the UCMs by considering  the effect not only of
$T_{\rm eff}$ and log\,$g$ but also of the change of the background 
opacities due to atoms, ions, dust grains, and quasi-continuous sources 
such as H$_2$ CIA \citep{tsu03}.

\subsection{The $J$-Band Region}

The spectra between 0.95 and 1.3 $\mu$m are shown in Fig.4. First, we
confirm that FeH $F^4\Delta - X^4\Delta$ 0.99 $\mu$m bands can be
clearly seen not only in L dwarfs but also in T dwarfs as noted by
\citet{bur02b}. The 0.99 $\mu$m FeH bands gradually weaken from L6.5 
to L8 and further to T2, but strengthen again at T3.5. The weakening
of the 0.99 $\mu$m FeH bands from L6.5 to L8 is consistent with the
weakening of the 1.61 and 1.63 $\mu$m features in the same objects,
and this lends a support for the contribution of FeH to these
features. However, the observed strengths of FeH 0.99 $\mu$m bands,
not only of T dwarfs but also of L dwarfs, are difficult to interpret
directly by our UCMs as well as by other models now available. In this 
connection, \citet{bur02b} suggested that the dust cloud breaks in the 
early T dwarfs and FeH formed deep in the photosphere can be peered 
through the holes of the cloud. The idea of cloud clearing model  was 
motivated by 
the difficulty of the cloudy models
by \citet{mar02} to explain the observed color-magnitude diagram, 
but we have shown that 
the color-magnitude diagram can be accounted for by our UCM without 
introducing such a hypothesis as break-up of the cloud \citep{tsu03a}.

As an alternative interpretation, we would like to call attention to
the possible formation of the second convective zone near the surface
of the late L and early T dwarfs. This second convective zone is
predicted by the UCMs of $T_{\rm eff} \approx 1200 - 1600$ K and is
formed by the steep temperature gradient due to the large opacity of
the dust cloud itself \citep{tsu02}. The lower boundary of this
convective zone is facing the region free of dust but FeH can be
abundant there. Then these FeH molecules will be dredged up by the
convection to the upper cooler region and some FeH molecules will
remain super-saturated until they are eventually transformed to the
condensates. In fact, the second convective zone is rather thin and
some FeH molecules will survive at the upper boundary of the
convective zone. These FeH molecules will constantly be replenished by 
the convection and can be observed so long as the second convective
zone reaches the region to give observable effect. 
Such a possibility of vertical transport of FeH was also noted by
\citet{bur02b} who, however, remarked that the fragility of the FeH bond     
(dissociation energy  of only 1.67\,eV) will preclude such a possibility, 
even if such a  vertical mixing 
may account for the unexpectedly strong CO fundamental bands observed 
in the T dwarf Gl 229B \citep{nol97,opp98}. Certainly, more detailed
quantitative analysis will required before we reach a definite conclusion.

The other prominent features are K I doublets at 1.1690/1.1773 and at
1.2432/1.2522 $\mu$m. The K I lines are rather strong in the early and 
middle L dwarfs, but weaken in the late L dwarfs. After passing the
minimum at L8, they are again reinforced in the early and middle T
dwarfs until masked by the strong 1.1 $\mu$m water bands in the late T 
dwarfs. These results are quite consistent with those reported by
\citet{bur02a}. The diminishing K I lines towards later L
dwarfs can be interpreted as due to the effect of extinction by dust
which is located in the optically thin photosphere in L dwarfs and 
shows the maximum at late L dwarfs. For this reason, K I lines show
the minimum at late L as the water bands at 1.1 and 1.4 $\mu$m. After the
cloud is buried below the observable photosphere, K I lines strengthen
again despite the unfavorable excitation at lower temperatures until
being masked by rapidly strengthened water bands.

The feature near 1.14 $\mu$m is identified as due to Na I doublet
1.1404/1.138 $\mu$m with log$gf$ = $-$0.186/$-$0.487 \citep{wie69}. 
However, the observed feature near 1.14 $\mu$m shows asymmetry with
the stronger absorption at the short wavelength side, contradicting
the expectation from the gf-values.  Thus, there should be some other
contributions such as of H$_2$O and CH$_4$  to the 1.14 $\mu$m feature 
which, however, shows the similar pattern as the K I lines.

The 1.1 $\mu$m bands of water cannot be seen in the L5 dwarf, SDSS2249+00,
and this is unusually weak for L5, compared with the published results 
for other L5 dwarfs (e.g. Geballe et al. 2002, Reid et al. 2001b). Also 
K I lines just discussed above are also unusually weak for L5 in
SDSS2249+00. Also remembering the unique spectrum around 1.7 $\mu$m as
noted in \S3.2, this object may be peculiar for an L5 dwarf. The 1.1
$\mu$m bands of water are stronger in the L6.5 dwarf, 2MASS1711+22 than in
the L8 dwarf, 2MASS1523+30 as already noted.

\section{BOLOMETRIC LUMINOSITIES AND EFFECTIVE TEMPERATURES}

Since the observed infrared spectrum  covers the major part of the
total flux, it can be a useful measure of the bolometric luminosity if 
combined with a parallax measurement.
Since the number of the objects with known parallaxes is modest in
our sample, we also apply 
$K$-band bolometric correction (BC$_K$) to the photometric
data of the objects with known parallaxes in the literature.
We then estimate effective temperatures based on the
bolometric luminosities and the radius which can be nearly constant at
the Jupiter radius.

\subsection{Objects with Known Parallaxes in Our Sample}

A direct measurement of the bolometric flux is difficult for brown 
dwarfs at present. However, a large part of the luminosity is found in 
the near infrared region we have observed, namely between 0.87 and 2.5 
$\mu$m. This is predicted by the use of  SEDs based on
UCMs \citep{tsu02} and  the fraction of the integrated flux
between 0.87 and 2.5 $\mu$m over the total flux is shown in Fig.5a.
In Fig.5a, three curves are given for $\log g$ = 4.5, 5.0, and 5.5 
($T_{\rm cr}$ = 1800 K).
Since they are not very different, we use the one for $\log g$ = 5.0.
It turns out that the near infrared flux is about 50\% of the total
flux at $T_{\rm eff} \approx 700$ K, but nearly 70\% at 
$T_{\rm eff} \approx 2000$ K. The inversion around $T_{\rm eff}
\approx 1600$ K is due to the dust extinction which is the largest around
that temperature. 
We integrate our observed spectra of the objects with known
parallaxes, which are flux calibrated
as noted in \S2 and then the bolometric fluxes are estimated by
applying the ratio of the integrated infrared flux to the bolometric
flux by the use of the model prediction shown in Fig.5a. 
To obtain $T_{\rm eff}$, we assume the range of the brown dwarf radius 
$R$
to be from  $1.0 R_J$ (relatively young object) to
$0.85 R_J$ (very old object) \citep{bur97}.
2MASS1146+22AB is assumed to be an equal binary composed of two L3
dwarfs, since the magnitude difference at $I$ is small \citep{rei01a}.
The resultant
bolometric luminosities and effective temperatures are given in Table 4.

An alternative approach to obtain bolometric luminosities is 
to apply bolometric correction to the absolute magnitudes.
The bolometric correction BC$_K$ for $K$-magnitude is given by

\begin{equation}
{\rm BC}_K = M_{\rm bol} - M_K,
\end{equation}

and this BC$_K$ has been evaluated by the use of UCMs in the CIT photometric
system.
The bolometric correction 
given as a function of $T_{\rm eff}$ is  shown
in Fig.5b, where three curves for $\log g$ = 4.5, 5.0, and 5.5 are
drawn. 
Again the gravity dependence is weak and we use the curve for $\log g$
= 5.0.
In order to estimate $T_{\rm eff}$ of an object whose $M_K$ is given,
the following iterative procedure is required. First, a tentative
$T_{\rm eff}$ is assumed and corresponding BC$_K$ 
is evaluated from Fig. 5b. From BC$_K$, $M_{\rm bol}$ is derived 
and, by assuming the radius of the object, $T_{\rm eff}$ is obtained.
The initial and resultant $T_{\rm eff}$s are compared. If they agree
to each other, the procedure is stopped. If they do not, the resultant
$T_{\rm eff}$ is used as the initial value and the procedure is repeated. 
Normally $T_{\rm eff}$ converges after a few iterations.
Again we assume the range of the radius to be from  $1.0 R_J$ to 0.85 $R_J$.
The bolometric luminosities 
and effective temperatures obtained this way are given in Table 4.

In Table 4, $L_{\rm bol}$(integrated flux) and 
$L_{\rm bol}$(BC$_K$) show reasonable agreement except for
2MASS1146+22A for which the difference is by a factor of 2.9.
As we find later, effective temperature estimates for 
early L dwarfs obtained by the 
$K$-band bolometric correction using UCMs tend to be higher than the estimates
by other methods. 
The agreement of the two methods for objects with
spectral types L5 and later is encouraging in that the $K$-band
bolometric correction may be applied to all objects with known
parallaxes to derive the relation between the effective temperature
and spectral type for a larger sample. 

\subsection{$T_{\rm eff}$ vs. Spectral Type}

Parallaxes and photometry  of L and T dwarfs have been obtained,
compiled and published 
in the literature  
(Burgasser 2001, Dahn et al. 2002; Tinney et al. 2003; references therein). 
Some authors have derived relations between $T_{\rm eff}$ and spectral
type for L dwarfs \citep{leg02,dah02} and for L and T dwarfs
\citep{bur01}, but there
has not been a work in which the relation was obtained based on
bolometric correction derived from photospheric models which are
applicable throughout the L-T sequence. In this subsection, we apply the
bolometric correction derived at $K$ band from UCMs to
obtain the relation between $T_{\rm eff}$ and spectral type.

39 parallax measurements of 36 objects
from \citet{bur01} (references therein), \citet{dah02}
and \citet{tin03} are used. 
For SDSS1254$-$01 and SDSS1624+00,
different values of the parallaxes have been obtained by Dahn et al.
and Tinney et al..
For a binary,  
Tinney et al. (references therein) estimate the contribution
to broadband photometry
and spectral type of each component. 
We adopt the estimate  of the $K$ magnitude and spectral type 
of the primary component by Tinney et al., 
but we do not include the secondary component. 
Since the spectral type of the
secondary component was not determined by spectroscopy, but by the magnitude
difference, it is not logical to use it to derive the relation
between $T_{\rm eff}$ and spectral type.
2MASS0559$-$14 appears exceedingly overluminous 
and the range of  $T_{\rm eff}$ is 
obtained also for the case of an equal binary
\citep{bur01}. \citet{bur03} obtained an HST image of this object
and did not resolve it into a binary with a resolution of 0.5 AU.
The resultant effective temperatures are given in Table 5.  

The relation between $T_{\rm eff}$ and 
the spectral type is plotted in Fig. 6.
Apart from the two overluminous objects, Kelu1 and Gl417B,
$T_{\rm eff}$ monotonically decreases toward later spectral types. 
$T_{\rm eff}$ appears too high for early L, but reasonable otherwise.
This problem may have some bearing on the fact that
a significant fraction of early L dwarfs are main-sequence stars.
Another possibility is that the simplest assumption we adopted
that the cloud properties do not change throughout the entire
L-T sequence may not
be valid.

\subsection{Some Representative Objects}

In this subsection, we discuss representative three objects
in comparison with works by others.

{\bf Gl229B}:
The first T dwarf, Gl229B \citep{nak95}, is also
the most extensively observed brown dwarf so far.
\citet{mat96} used broadband photometry from 0.8 to 10.5
$\mu$m to derive the observed luminosity in the atmospheric windows,  
and a dust-free model \citep{tsu96} to derive the
bolometric luminosity, 6.4$\times 10^{-6} L_\odot$. Assuming the
radius of Gl229B to be $R_J$, they gave the best estimate of 
the effective temperature,  
$T_{\rm eff}$ = 900K.
Leggett et al. (1999) used $JHK$ spectra and $L$-band photometry,
and a model \citep{all01} to estimate the bolometric
luminosity,   6.6$\times 10^{-6} L_\odot$. From a comparison with
the evolutionary model by Burrows et al. (1997), they also gave
an estimate of $T_{\rm eff}$ = 900K.
In this paper, we obtained the luminosity
$L_{\rm bol}$(integrated flux)=$6.02\times10^{-6} L_\odot$ 
and the effective temperature,
$T_{\rm eff}$(integrated flux)=905K for $R=R_J$. 
Since UCMs become indistinguishable from dust-free models for the late T
dwarfs, this result is as expected and $T_{\rm eff}$s for the late T dwarfs
are determined well.

{\bf SDSS1254$-$01}:
This is one of the first L/T transition objects discovered by 
SDSS \citep{leg00}. Determination of its effective temperature 
directly leads to that of the L/T transition temperature.
\citet{bur01} estimated the effective temperature of this object
by linear interpolation of the relation between effective temperature
and spectral type obtained empirically to be $T_{\rm eff}=1270\pm120$K. 
\citet{dah02} and \citet{tin03} obtained significantly
different values for the parallax and we calculated 
$L_{\rm bol}$(integrated flux)=$2.20\times10^{-5} L_\odot$ 
for the parallax measurement by Dahn et al. and 
$L_{\rm bol}$(integrated flux)=$2.96\times10^{-5} L_\odot$ 
for that by Tinney et al..
Corresponding effective temperatures are 
$T_{\rm eff}$=1252K (parallax from Dahn et al.) and  
$T_{\rm eff}$=1348K (parallax from Tinney et al.) for $R=R_J$.
In the separate paper \citep{tsu03}, we show that 
the observed spectrum of SDSS1254$-$01 is fitted well by
the UCM model spectrum for $T_{\rm eff}$=1300K. 
The agreement of the $T_{\rm eff}$s obtained by the different methods
indicates that the L/T transition temperature is well constrained to about 1300K.

{\bf 2MASS1523+30}:
This is one of the latest L dwarfs discovered by 2MASS \citep{kir00}
and is also known as Gl 584C \citet{kir01}, a companion to a G star.
\citet{kir00}, \citet{bur01}, and \citet{dah02} used 
empirical bolometric correction to estimate $T_{\rm eff}\approx 1300$K,
$T_{\rm eff}=1240\pm80$K,
and $T_{\rm eff}=1376\pm58$K respectively. 
\citet{leg02} obtained the bolometric luminosity of this
object by using the integrated flux from the red to $K$, $L^\prime$
photometry, and assuming a Rayleigh-Jeans curve longward of $L^\prime$
to be $2.69\times10^{-5} L_\odot$. They used the evolutionary models 
of \citet{cha00} to estimate the range of effective
temperature, $T_{\rm eff} =1250-1500$K.  
We obtained the bolometric luminosity, $2.46\times10^{-5} L_\odot$,
which is consistent with the one by Leggett et al..
We gave the range of effective temperature to be $T_{\rm eff} =1252-1358$K.   
So $T_{\rm eff}$ of this object is not significantly different from
that of SDSS1254$-$01. 
From an analysis of effective temperatures obtained by
empirical bolometric correction, \citet{bur01} noted that there is
essentially no change in effective temperatures between types L8V and
T5V. Here we confirm his observation for the range between L8 and T2.
The relation between the effective temperature and spectral type
derived by the $K$-band bolometric correction obtained above
further indicates that
effective temperatures do not change significantly between L7 and T5.

\subsection{$T_{\rm eff}$-Spectral Type Relations Obtained by Others}

\citet{ste01} obtained the trends observed in the ($K-L^\prime$) and 
($K-L_s$) colors as a function of L and T dwarf spectral class.
They compared these colors with theoretical models \citep{mar02}, 
derived a relationship
between $T_{\rm eff}$ and L spectral class:

\begin{equation}
T_{\rm eff} = 2200 - 100 L_K,
\end{equation}

where $L_K$ is the spectral class by \citet{kir99}, which ranges from 0 to 8.
They noted that the equation was presented for its heuristic value as an
indicator of the $T_{\rm eff}$ range of L dwarfs and not as a definite 
analysis. In the narrow range between L3 and L8, this linear relation appears 
consistent with the relation we obtained in the previous section.

\citet{leg02} determined the luminosities of 2M0036+18 (L4), 2M0825+21 (L6),
2M1523 (L8), and 2M1632+19 (L7.5), by summing their energy distributions
from the red to the $K$ band, interpolating the flux between the
$K$ band and the effective $L^\prime$ flux computed from their photometry,
and assuming Rayleigh-Jeans curves longward of $L^\prime$. 
A correction to this simple approach was determined for Gl 229B by
\citet{leg99} using model atmospheres. \citet{leg99} also used
models to show that no correction is needed for dwarfs as late as 
mid L. \citet{leg02} adopted a correction for the 2MASS L7.5 and L8
dwarfs that is half that computed for Gl 229B.
They compiled the bolometric luminosities of 18 dwarfs and used
the models of \citet{cha00} to compute $T_{\rm eff}$ for the L dwarfs
for which they obtained the bolometric luminosities.
Their compilation of $T_{\rm eff}$ for L dwarfs is consistent with
the our result between L3 and L8 within given error estimates.
For early L dwarfs, our temperature estimates are somewhat higher.
Their sample does not include early T dwarfs, but they estimated
that the effective temperature range for the L dwarfs in their
sample was approximately 2200-1300K and for the T dwarfs 1300-800K.
Their estimate of the L/T transition temperature, 1300K is in agreement
with our result.

\citet{dah02} and \citet{bur01} 
used empirical bolometric corrections of \citet{leg02}
and \citet{rei01b} to derive $T_{\rm  eff}$ and \citet{bur01}
obtained that for T dwarfs by linear approximation.
The values of $T_{\rm eff}$ of L dwarfs obtained by \citet{dah02} are mostly
consistent with our estimates. \citet{bur01} linearly fitted
the compiled data of L dwarfs to obtain the relation:

\begin{equation}
T_{\rm eff} = (2380\pm40) - (138\pm8) SpT
\end{equation}
where $SpT$ ranges from L0 to L8. 
This relation shows somewhat better agreement with our data
than that by \citet{ste01} between L0 and L8.

\citet{bur02b} adopt a completely different approach
in estimating the effective temperatures of L/T transition objects.
They consider that a cloudy model alone cannot explain the 
L/T transition process due to the failure of the cloudy model by
\citet{mar02} in reproducing the color-magnitude diagram.
In their cloud clearing model, they assume that the location of
the L/T transition object in the color-magnitude
diagram is determined by the fraction of the
clear region on the surface of the brown dwarf and it is compared
with the predicted color-magnitude by the mixture of the cloudy
model and the cloud free model. Their estimate of the L/T transition
temperature is 1200K, which is about 100K cooler than ours.
Here we briefly examine the cloud clearing model.
First, one underlying assumption is that
the model by \citet{mar02} is correct as a cloudy model, about which
we have some doubt. UCMs \citep{tsu03a} reproduced the basic features
of the 
color-magnitude diagram of L and T dwarfs 
although the basic assumptions on the clouds are simpler. 
Second, the cloud clearing model has not passed the test of
comparing the spectra of L/T transition objects
with model spectra. Third, \citet{eno03} observed variabilities
of L and T dwarfs and did not find significant correlations between
variability amplitude and spectral type or $J-K$ color. This result does
not support the cloud clearing model positively.  
\citet{bur02b} claim that the plateau of the effective temperature
near the L/T transition, which we also see
in Fig. 6  is naturally explained by the cloud clearing
model, but there is a 100K difference in the transition temperature.

In summary, UCM estimates of L dwarf temperatures are consistent with
others except for early L for which the UCMs give somewhat higher
temperatures. The estimate of the L/T transition temperature, 1300K,
is in agreement with that by \citet{leg02}, but somewhat higher than
that by \citet{bur02b}. 

\section{SUMMARY AND CONCLUDING REMARKS}

We have obtained near-infrared
spectra of L dwarfs, L/T transition objects
and T dwarfs using Subaru.
The resulting spectra are examined in detail to see their dependence 
on the spectral types. 
As for methane, we have found  that it appears at L8 and marginally at L6.5.
The water bands at 1.1 and 1.4 $\mu$m do not necessarily show steady
increase towards later L types but may show inversion at late L types.
This does not necessarily imply that the spectral types do not 
represent a temperature sequence, but can be interpreted as due to
compensation of the increasing water abundance by the dust extinction
in the later L types. We confirm that the FeH 0.99 $\mu$m bands appear 
not only in late L dwarfs but also in the early T dwarfs
as was first noted by \citet{bur02b}. We suggest
that FeH could be dredged up by the surface convective zone induced by 
the steep temperature gradient due to the large opacity of the dust
cloud itself and will be replenished constantly by the convection.

We have obtained bolometric luminosities of the objects with known
parallaxes in our sample
by integrating the spectra between 0.87 and 2.5 $\mu$m
and by the $K$-band bolometric correction. 
Apart from the L3 dwarf, 2MASS1146+22A, the bolometric luminosities obtained by
both methods agree well and this implies that the $K$-band 
bolometric correction obtained by UCMs
can be used to obtain the bolometric luminosities and
effective temperatures of the L and T dwarfs
with known parallaxes from the literature. 
The relation between the effective temperature and spectral type
derived from the $K$-band bolometric correction
shows  monotonic 
behavior throughout the L-T sequence. 

There is another method to estimate the effective temperature of a
brown dwarf, which is to compare the observed spectrum and model
spectra.
Such an analysis is also vital as a test of model photospheres.
For these purposes, we analyze the observed spectra discussed in this
paper
with the model spectra obtained by an extended grid of UCMs in the
separate paper \citep{tsu03}.




\acknowledgments

We thank the support astronomer K. Aoki and the staff of the Subaru
Observatory for the excellent support of the observations.
We also thank the referee, A. Burgasser for his careful reading 
of the manuscript.
This work was supported by the grants-in-aid of JSPS Nos. 14520232
(T.N.) and 11640227 (T.T.).

\clearpage

\begin{deluxetable}{cccccccc}
\footnotesize
\tablecaption{Photometry \label{tbl-1}}
\tablewidth{0pt}
\tablehead{
\colhead{Object} & \colhead{RA} & \colhead{DEC} &
\colhead{$J$}   & \colhead{$H$}   &  \colhead{$K^\prime$} 
& \colhead{$K$} & \colhead{Source} 
} 
\startdata
2MASS1217$-$03 & 121711.0 & $-$031113 &  15.41 & 15.96 & 15.62 & & this work \nl
SDSS1254$-$01 & 125453.9 & $-$012247 &  14.54 & 14.03 & 13.82 & & this work \nl
2MASS1523+30 & 152322.6 & +301456 &  15.93 & 15.04 & 14.44 & & this work \nl
2MASS1711+22 & 171145.7 & +223204 &  16.57 & 15.86 & 15.20 & & this work \nl
SDSS1750+17 & 175033.0 & +175904 &  16.14  & 15.94  & & 16.02  & Leggett et
al. 2002 \nl
SDSS2249+00 & 224953.5 & +004404 &  16.46  & 15.42  & & 14.43  & Leggett et
al. 2002 \nl
\enddata

\tablenotetext{a}{Photometric uncertainties are 0.05 mag}

\end{deluxetable}

\clearpage

\begin{deluxetable}{cccc}
\footnotesize
\tablecaption{Spectroscopy \label{tbl-2}}
\tablewidth{0pt}
\tablehead{
\colhead{Object} & \colhead{Grism}   & \colhead{Integration time}   &  
\colhead{Reference star}  
} 
\startdata
2MASS1217$-$03 & zJ & 200s $\times$ 4 & SAO138672 (G0) \nl
       & JH & 200s $\times$ 4 & \nl
       & wK & 200s $\times$ 4 & \nl
SDSS1254$-$01 & zJ & 100s $\times$ 4 & SAO139015 (G0) \nl
       & JH &  50s $\times$ 4 & \nl
       & wK &  50s $\times$ 4 & \nl
2MASS1523+30 & zJ & 200s $\times$ 4 & SAO64641 (GO) \nl
       & JH & 100s $\times$ 4 & \nl
       & wK & 100s $\times$ 4 & \nl
2MASS1711+22 & zJ & 400s $\times$ 4 & SAO84938 (F8) \nl
       & JH & 100s $\times$ 8 & \nl
       & wK & 100s $\times$ 4 & \nl
SDSS1750+17 & zJ & 200s $\times$ 4 & SAO85439 (G0) \nl
       & JH & 100s $\times$ 8 & \nl
       & wK & 200s $\times$ 4 & \nl
SDSS2249+00 & JH & 200s $\times$ 4 & SAO127766 (F8) \nl

\enddata

 
 
\end{deluxetable}

\clearpage

\begin{deluxetable}{cccc}
\rotate
\tablecaption{Sample for Spectral Analysis}
\tablewidth{0pt}
\tablehead{
\colhead{Object} & \colhead{Spectral Type} & \colhead{Source} & 
\colhead{Discovery}\\
}
\startdata
2MASS1146+22AB & L3 & Nakajima et al. 2001 & Kirkpatrick et al. 1999\nl
2MASS1507$-$16   & L5 & Nakajima et al. 2001 & Kirkpatrick et al. 2000\nl
SDSS2249+00   & L5\tablenotemark{a} & this work & Geballe et al. 2002\nl
2MASS1711+22   & L6.5 & this work & Kirkpatrick et al. 2000\nl
2MASS0920+35   & L6.5 & Nakajima et al. 2001 & Kirkpatrick et al. 2000\nl
2MASS1523+30   & L8   & this work & Kirkpatrick et al. 2000\nl
SDSS1254$-$01   & T2\tablenotemark{a}   & this work & Leggett et al. 2000 \nl
SDSS1750+17   & T3\tablenotemark{a}   & this work & Geballe et al. 2002 \nl
Gl 229B  & T6\tablenotemark{a} 
& Geballe et al. 1996, Leggett et al. 1999 & Nakajima
et al. 1995 \nl
2MASS1217$-$03   & T7.5\tablenotemark{b} & this work & Burgasser et al. 1999\nl
\enddata

\tablenotetext{a}{Infrared classification by \citep{geb02}}
\tablenotetext{b}{Infrared classification by \citep{bur02a}}

\end{deluxetable}

\clearpage 

\begin{deluxetable}{cccccc}
\rotate
\footnotesize
\tablecaption{Bolometric Luminosities \& Effective Temperatures \label{tbl-4}}
\tablewidth{0pt}
\tablehead{
\colhead{Object} & Spectral Type & \colhead{$L_{\rm bol}$(integrated flux)} & 
\colhead{$T_{\rm eff}$(integrated flux)}   
& \colhead{$L_{\rm bol}$(BC$_K$)} & \colhead{$T_{\rm eff}$(BC$_K$)}
\\
 & & $L_\odot$ & K & $L_\odot$ & K 
} 
\startdata
2MASS1146+22A & L3 & 6.06$\times 10^{-4}$ & 1612-1748 & 1.73$\times 10^{-4}$ & 2098-2276 \nl
2MASS1507$-$16  & L5 & 3.17$\times 10^{-5}$ &1371-1487  & 5.51$\times 10^{-5}$ & 1544-1675 \nl
2MASS1523+30 & L8 & 2.46$\times 10^{-5}$ & 1287-1395   & 3.63$\times 10^{-5}$ & 1418-1538\nl
SDSS1254$-$01 \tablenotemark{a} 
& T2 & 2.20$\times 10^{-5}$ & 1252-1358  &  2.46$\times10^{-5}$ & 1286-1395\nl 
SDSS1254$-$01 \tablenotemark{b}
&    & 2.96$\times 10^{-5}$ & 1348-1462  &  2.91$\times10^{-5}$ & 1342-1456\nl 
Gl229B & T6.5 &6.02$\times 10^{-6}$ & 905-981  & 6.67$\times10^{-6}$  & 928-1007 \nl 
2MASS1217$-$03 & T7.5 &5.50$\times 10^{-6}$ & 885-960  &        5.22$\times10^{-6}$  & 873-947 \nl
\enddata

\tablenotetext{a}{Parallax data from Dahn et al. (2002)}
\tablenotetext{b}{Parallax data from Tinney et al. (2003)}

\end{deluxetable}

\clearpage

\begin{table*}

\begin{center}

Table 5. Spectral Type vs. Effective Temperature

\footnotesize
\begin{tabular}{cccccccc}
\tableline
Object    &   Sp. Type & $M_K$ 
&  $M_{\rm bol}$ & BC$_K$ & $T_{\rm min}$ & $T_{\rm max}$ 
& Refs \tablenotemark{a} \\
          &             &                &        & K & K 
& & \\
\tableline
HD89744B   &  L0 &  10.65 &   13.56& 2.91& 2398& 2601 & K00,D \\
2MASS0345+25  &  L0 & 10.58 & 13.49& 2.91& 2437& 2643 & K99,D \\
2MASS0746+20A &  L0.5 & 10.65 &   13.56& 2.91& 2398& 2601& K00,D,T \\
2MASS1439+19  &  L1 & 10.08 &  13.71& 2.91& 2316& 2512 & K99,D\\
2MASS1658+70  &  L1 & 10.57 &    13.48& 2.91& 2442& 2649 & K00,D\\
GJ1048B    &  L1 & 10.67 &   13.58& 2.91& 2387& 2589 & GKW,P,D \\
Kelu1      &  L2 & 10.46 &   13.37& 2.91& 2505& 2717 & D,K99,K00\\
Gl618.1B   &  L2.5& 11.18 &  14.20& 3.02& 2069& 2244 & W,P,B\\
DENIS1058$-$05  &  L3& 11.48 &    14.72& 3.24& 1836& 1991 & K99,K00\\
2MASS1146+22A &  L3&  11.12 &   14.14& 3.02& 2098& 2276 & K99,D,T \\
2MASS0036+18  &  L3.5& 11.38 &   14.62& 3.24& 1878& 2037 & K99,D\\
2MASS0326+29  &  L3.5& 11.09 &   14.22& 3.13& 2060& 2234 & K99,D\\
GD165B     &  L4 & 11.52 &   14.89& 3.37& 1765& 1915 & K99,A\\
2MASS2224$-$01  &  L4.5& 11.76 &  15.20& 3.44& 1643& 1783 & K00,D\\
Gl417B     &  L4.5& 11.01 &  13.97& 2.96& 2182& 2366 & K01,P \\
2MASS1507$-$16  &  L5 & 12.07 &   15.47& 3.40& 1544& 1675 & K00,D\\
GJ1001B    &  L5 & 11.50 &   14.87& 3.37& 1773& 1923 &K01,A\\
DENIS1228$-$15A&  L5 & 12.00 &    15.40& 3.40& 1567& 1702 &K99,D,T\\
2MASS1328+21  &  L5 & 11.64 &   15.08& 3.44& 1690& 1833 &K99,K00\\
2MASS0850+10A&  L6 &  12.44 &  15.84& 3.40& 1418& 1538 &K99,D\\
2MASS0825+21  &  L7 & 12.99 &   15.91& 2.92& 1396& 1514 &K99,D\\
DENIS0205$-$11A &  L7 & 12.28 &  15.53& 3.25& 1523& 1652 &K99,T\\
2MASS1632+19  &  L8 & 13.04 &   15.96& 2.92& 1418& 1538 &K99,D\\
2MASS1523+30,Gl584C     &  L8 & 12.92 &  15.84& 2.92& 1418& 1538 &K01,P,D\\
Gl337C     &  L8 &  12.46 & 15.71& 3.25& 1461& 1585 & W,P\\
SDSS1254$-$01  &  T2 & 13.48 &   16.26& 2.78& 1286& 1395 &L02,D \\
SDSS1254$-$01     &  T2 & 13.16 &  16.08& 2.92& 1341& 1455 &L02,T\\
SDSS1021$-$03     &  T3 & 12.94 &   15.96& 3.02& 1380& 1496 &B,T,L02\\
2MASS0559$-$14  &  T5 & 13.58 &   16.21& 2.63& 1302& 1413 &B,D\\
2MASS0554$-$14A &  T5 & 14.33 &   16.74& 2.41& 1153& 1250 &B,D \\
2MASS1534$-$29A     &  T5.5& 15.00 &  17.24& 2.24& 1027& 1114 &B,T\\
SDSS1624+00  &  T6 & 15.42 &   17.49& 2.07&  970& 1052 &L02,D\\
SDSS1624+00     &  T6 & 15.40 &   17.47& 2.07&  974& 1057 &L02,T\\
2MASS1225$-$29A     &  T6 & 14.82 &   17.06& 2.24& 1071& 1161 &B,T,L02\\
2MASS1346$-$00     &  T6 & 14.90 &   17.13& 2.23& 1054& 1143 &B,T,L02\\
Gl229B     &  T6.5& 15.61 &   17.68& 2.07&  928& 1007 & B,L99,P\\
2MASS1047+21     &  T6.5& 16.42 &  18.34& 1.92&  797&  865 & B,T,L02\\
2MASS1217$-$03     &  T7.5& 15.71 &  17.80& 2.09&  903&  979 &B,T,L02\\
Gl570D     &  T8 & 16.66 &   18.68& 1.51&  737&  800 &B,P,G\\
\tableline
\end{tabular}
\end{center}

\tablenotetext{a}{A:\citet{alt95}; B:\citet{bur01}; D:\citet{dah02}; G:\citet{geb01}; GKW:\citet{giz01};K99:\citet{kir99}; K00:\citet{kir00}; K01:\citet{kir01};L99:\citet{leg99}; L02:\citet{leg02}; P:\citet{per97};T:\citet{tin03}; W:\citet{wil01}}

\end{table*}

%
%
%



\clearpage

\clearpage

\figcaption[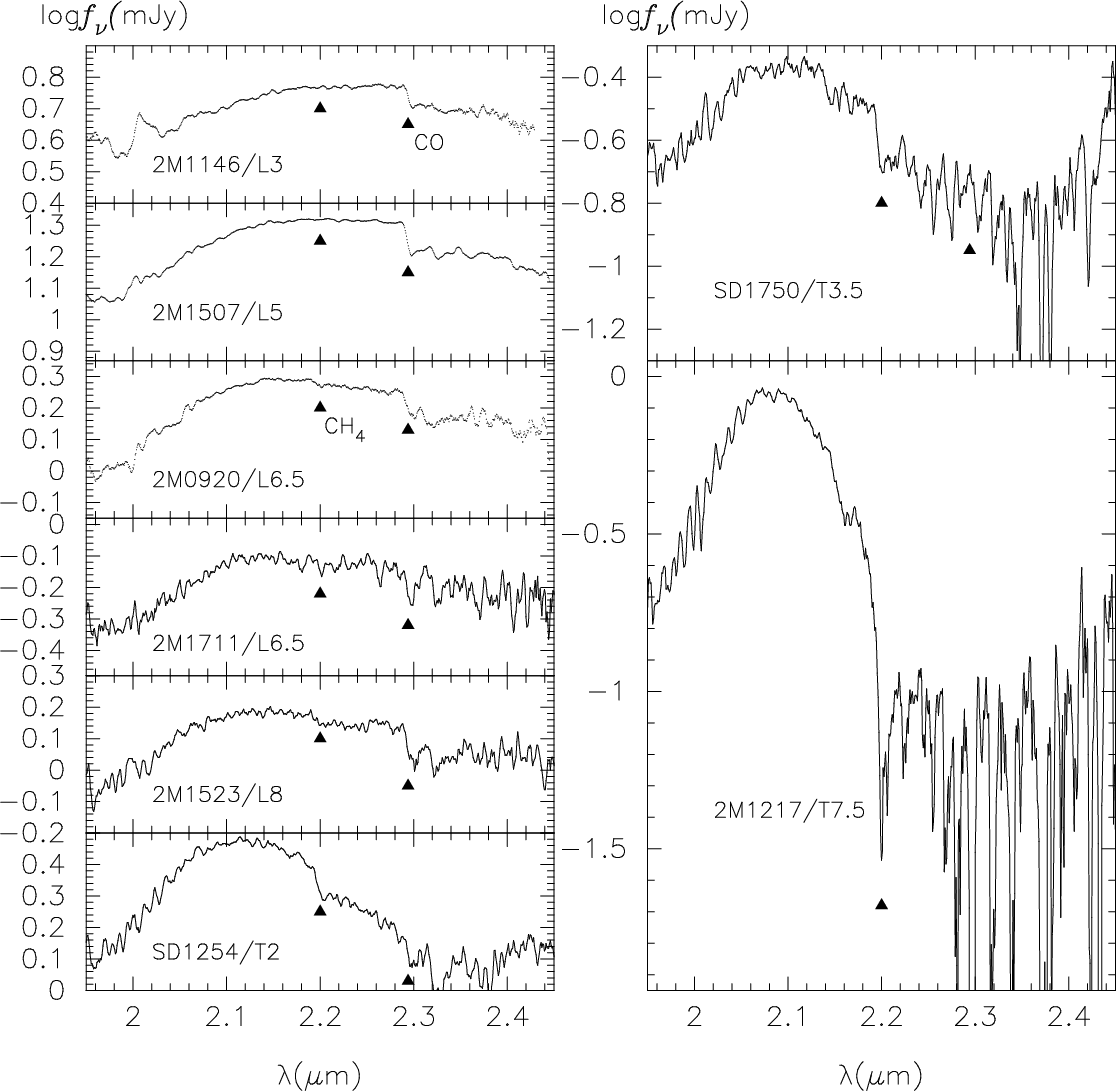]{$K$-band spectra. Spectra of eight objects
are shown on the $\log f_\nu$ scale. Filled triangles 
indicate CH$_4$ (2.2 $\mu$m) and 
CO (2.3 $\mu$m).  \label{fig1}}

\figcaption[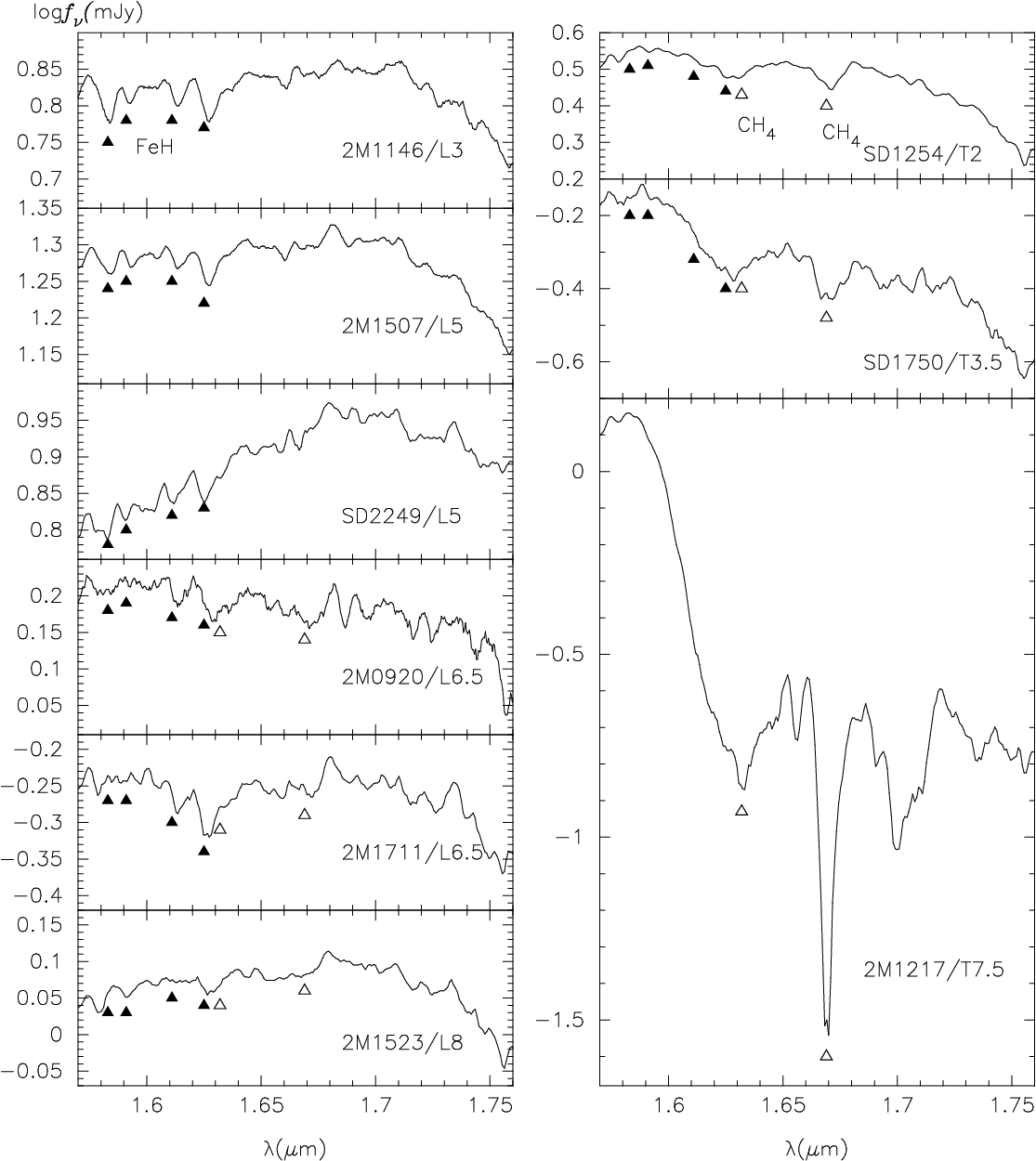]{ $H$-band spectra. Spectra of eight objects
are shown. Filled
triangles are the bandheads of FeH and open triangles are locations of 
CH$_4$. Note that the ordinate scales of the left and right
hand panels differ by factor of 2. \label{fig2}}

\figcaption[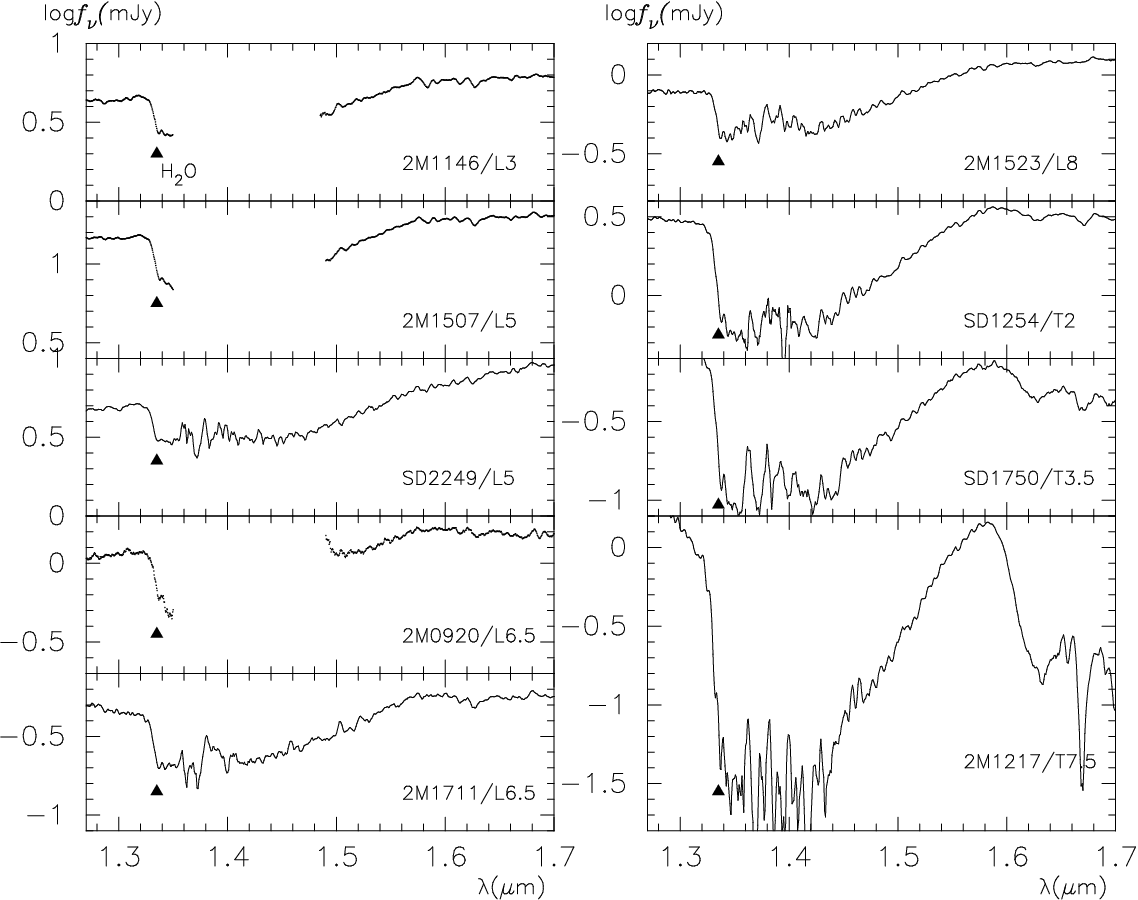]{Spectra between $J$ and $H$ bands.
Spectra of nine objects are shown. For the six objects
observed by CISCO (this work), the wavelength coverage is
continuous. Filled triangle indicates the location of 
H$_2$O. \label{fig3}}

\figcaption[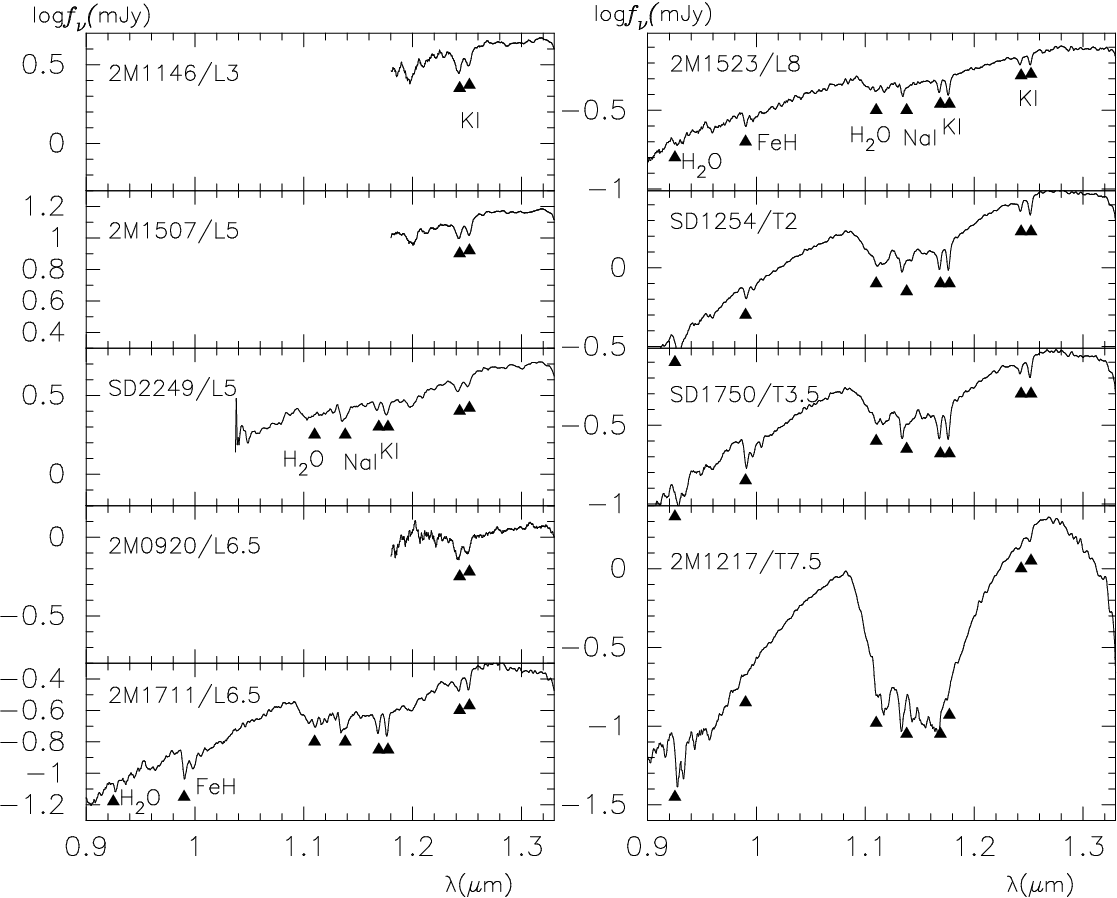]{$J$-band spectra. Spectra of nine objects
are shown. Filled triangles indicate FeH (0.99 $\mu$m),
H$_2$O (0.925 and 1.1 $\mu$m), Na I (1.14 $\mu$m), and K I (1.169/1.177 and 
1.243/1.252 $\mu$m). \label{fig4}}

\figcaption[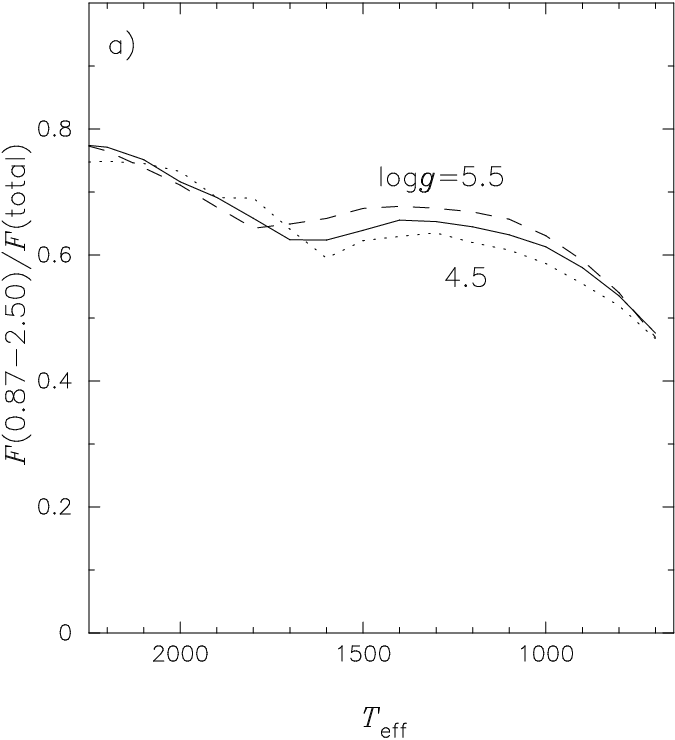,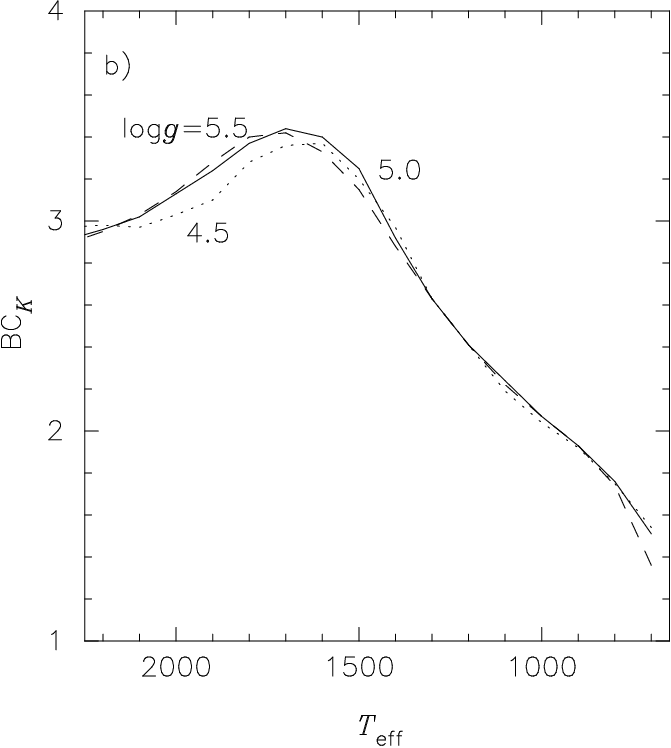]{(a) Fraction of integrated 
flux between 0.87 and
2.5 $\mu$m over total flux given as a function of $T_{\rm eff}$
estimated by the Unified Cloudy Models ($T_{\rm cr}$ = 1800 K). 
Three curves are drawn for
$\log g$ = 4.5 (dot), 5.0 (solid), and 5.5 (dash). The gravity dependence
is weak and we adopt the one for $\log g$ = 5.0.
(b) $K$-band bolometric correction given as a function of   $T_{\rm eff}$.
Three curves are drawn for $\log g$ = 4.5 (dot), 5.0 (solid), and 5.5 (dash).
The gravity dependence is weak and we adopt the one for $\log g$=5.0.}

\figcaption[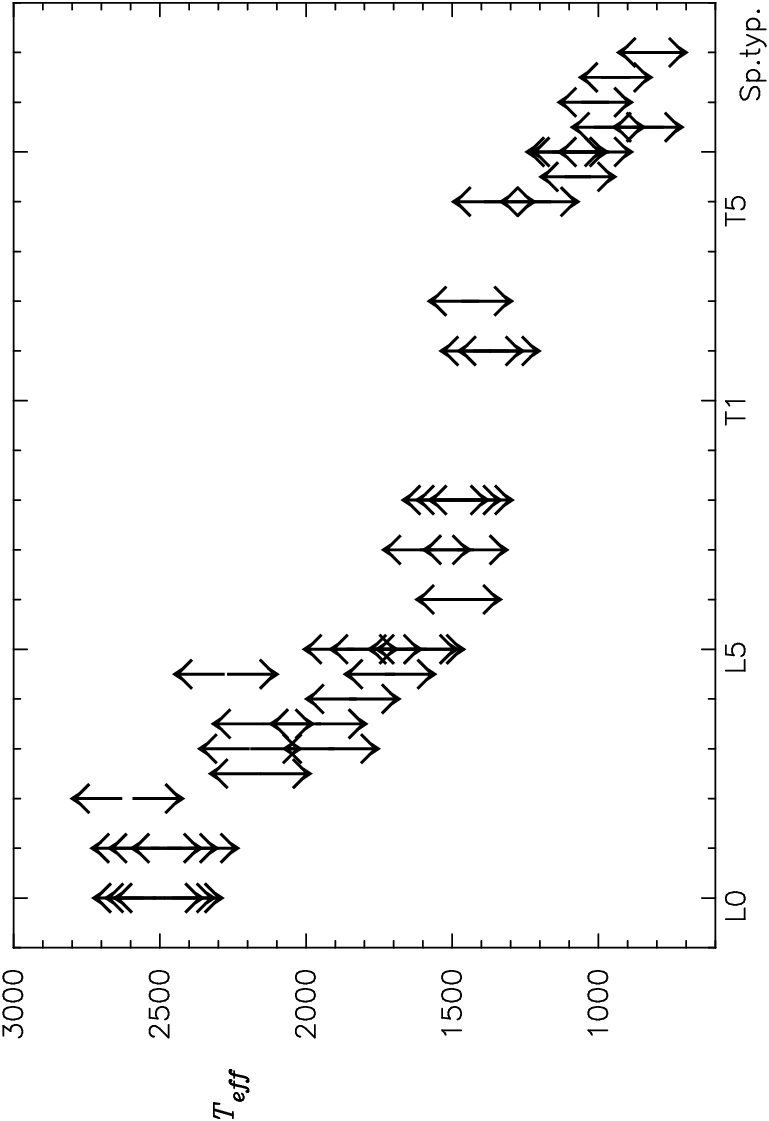]{Effective temperature vs. spectral type.
The higher and lower estimates
of $T_{\rm eff}$ obtained by the $K$-band bolometric correction 
are plotted as a function of spectral type.\label{fig6}}



\clearpage

\plotone{f1.eps}

\clearpage

\plotone{f2.eps}

\clearpage

\plotone{f3.eps}

\clearpage

\plotone{f4.eps}

\clearpage

\plotone{f5a.eps}

\clearpage

\plotone{f5b.eps}

\clearpage

\plotone{f6.eps}

\end{document}